\documentclass[aps,prl,10pt,twocolumn,superscriptaddress,showpacs]{revtex4-1}
\usepackage[dvipdfm]{graphicx}
\usepackage{natbib}
\usepackage{subfigure}
\usepackage{tabularx}
\usepackage{epsfig}
\usepackage{longtable}
\usepackage{amsfonts}
\usepackage{rotating}

\begin{document}

\title{Anomalous Nuclear Quantum Effects in Ice}

\author {B. Pamuk}
\affiliation{Department of Physics and Astronomy, Stony Brook University, Stony Brook, New York 11794-3800, USA}
\author{J. M. Soler}
\affiliation{Dep. de F\'{\i}sica de la Materia Condensada,
                        Universidad Aut\'onoma de Madrid, 28049 Madrid, Spain}
\author{R. Ram\'{i}rez}
\affiliation{Instituto de Ciencia de Materiales, Consejo Superior de Investigaciones Cientificas (CSIC), Campus de Cantoblanco, 28049 Madrid, Spain}
\author{C. P. Herrero}
\affiliation{Instituto de Ciencia de Materiales, Consejo Superior de Investigaciones Cientificas (CSIC), Campus de Cantoblanco, 28049 Madrid, Spain}
\author{P. W. Stephens}
\affiliation{Department of Physics and Astronomy,  Stony Brook University, Stony Brook, New York 11794-3800, USA}
\affiliation{Photon Sciences Directorate, Brookhaven National Lab. Upton, NY 11973, USA}
\author{P. B. Allen}
\affiliation{Department of Physics and Astronomy,  Stony Brook University, Stony Brook, New York 11794-3800, USA}
\author{M.-V. Fern\'andez-Serra}
\email{maria.fernandez-serra@stonybrook.edu}
\affiliation{Department of Physics and Astronomy,  Stony Brook University, Stony Brook, New York 11794-3800, USA}

\date{\today}

\begin{abstract}

	One striking anomaly of water ice has been largely neglected and never explained.  
	Replacing hydrogen ($^1$H) by deuterium ($^2$H) causes ice  to expand, whereas the ``normal"
	isotope effect is volume contraction with increased mass. 
	Furthermore, the anomaly increases with temperature $T$,
	even though a normal isotope shift should decrease with $T$ and vanish when $T$ is high enough to use classical nuclear motions. 
	In this study, we show that these effects are very well described 
	by {\it ab initio} density functional theory. 
	Our theoretical modeling explains these anomalies,  
	and allows us to predict  and to experimentally confirm a counter effect, namely that replacement of $^{16}$O by $^{18}$O
	causes a normal lattice contraction.


\end{abstract}

\pacs{31.15.A-,31.70.Ks, 65.40.De, 71.15.Pd}

\maketitle

Numerous recent studies~\cite{Soper08,Manolopoulos09,Ramirez2010a,Car08,michaelides11,Ramirez11,Zeidler11} 
address the contribution of zero-point nuclear
quantum effects to the structures of ice and water.
The issue is delicate, because of the peculiar electrostatic-covalent 
nature~\cite{mvfsprl06} of the hydrogen bond (Hbond) in water.
It is well known that hydrogen bonded materials show an 
anti-correlation~\cite{Libowitzky99} effect between the 
OH covalent bond and the OH--O Hbond.
As the OH-O distance diminishes, the OH covalent bond weakens 
(its length increases and vibrational frequency diminishes~\cite{Wilkinson84,Libnau94}) while the OH–-O Hbond does the opposite.
Closely related is the surprising expansion
of the ice (Ih) crystal~\cite{Kuhs94,Kuhs12}
 when H$_2$O is replaced by D$_2$O.
A similar effect is observed in water, where the molecular volume of 
H$_2$O is smaller than that of D$_2$O\cite{Kell67} at all temperatures.
Zero point expansion of crystal lattices is a well understood phenomenon,
 and is almost always greatest for the lightest isotopes.  
For example, the volume of $^{20}$Ne expands by 0.6\% with 
respect to $^{22}$Ne at $T$=0~\cite{Batchelder68}.
This ``normal" isotope effect corresponds to a $\sim$12\% zero-point expansion 
of $^{20}$Ne relative to a hypothetical ``classical" or ``frozen" lattice~\cite{Herrero01,Allen94}. 
 Since H$_2$O and Ne have similar molecular masses, one might expect similar effects.  
However, the volume of H$_2$O at $T = 0$ is
$\sim$0.1\% smaller than that of D$_2$O~\cite{Kuhs94,Kuhs12}.  
It has rarely been mentioned in the literature that this is opposite to the usual behavior, 
and no explanation has been offered. 

 In this paper, we explain this effect as an interesting coupling 
between quantum nuclear motion and hydrogen bonding, that may be relevant also to the structure of liquid water.  
 Our analysis shows that, despite the anomalous isotope effect,
 quantum ice actually has a volume 1\% larger 
 than it would have with classical nuclei. 
The effects are smaller than in Ne mostly because of delicate cancellations. 
We exploit these cancellations to make critical comparisons of:
(i) quasiharmonic theory versus fully anharmonic path-integral molecular dynamics (PIMD);
(ii) ab initio forces versus flexible and polarizable empirical force fields (EFF); 
and (iii) various flavors of ab initio density-functional theory (DFT) 
exchange and correlation (XC) density functionals (DF)
with and without inclusion of van der Waals (vdW) interactions.
  We find: (i) quasiharmonic theory is satisfactory for this problem; (ii) present state of the art EFF’s are not good enough to
describe nuclear quantum effects in water; 
and (iii) all the DFs considered describe qualitatively the anomalous effects,
although some versions perform better than others.

Within the volume-dependent quasiharmonic approximation (QHA), 
the equilibrium volume $V(T)$ is obtained by minimizing at each $T$
the Helmholtz free energy $F(V,T)$ ~\cite{Ziman,GapIce}:
\begin{eqnarray}
F(V,T) 
  &=& E_0(V) + \nonumber \\
  && \sum_k \left[ \frac{\hbar\omega_k(V) }{2} + 
            k_B T \ln \left(1-e^{-\hbar \omega_k(V) / k_B T}\right) \right]
\label{eq:quasi-harm}
\end{eqnarray}
where $E_0(V)$ is the energy for classical ($T=0$ or frozen) nuclei, at the relaxed 
atomic coordinates for each volume.
$\omega_k$ are the phonon frequencies, with $k$ combining the 
branch index and the phonon wave vector within the Brillouin zone.
Their volume dependence is linearized as:
\begin{equation}
\omega_k(V) = \omega_k(V_0) \left[1 - \frac{(V-V_0)}{V_0}\gamma_k\right]
\label{eq:omega-V}
\end{equation}
where $\gamma_k=- (V_0/\omega_k) (\partial \omega_k / \partial V)_{V_0}$
 is the Gr\"{u}neisen parameter of the mode,
and $V_0$ is the equilibrium volume of $E_0(V)$. 
$\omega_k(V_0)$  and $\gamma_k(V_0)$ are obtained by diagonalizing 
the dynamical matrix, computed by finite differences from the atomic
forces in a $(3 \times 3 \times 3)$ supercell,  at two volumes 
slightly below and above $V_0$.
As shown in the supplementary information~\cite{SI} (SI), this linearization 
is an excellent approximation to the full QHA, 
in which the frequencies are calculated at each volume.
It can be easily shown~\cite{SI} that at $T$=0 the quasi-harmonic volume shift
with respect to the frozen lattice is isotope dependent, and proportional
to the average  $\langle \gamma_k \omega_k \rangle$.
For increasing temperature, quantum effects become less relevant,
and the shift converges to the classical
result, where the change in volume is $\propto T$ and isotope independent.
Therefore, for high enough $T$, the differences between isotopes must vanish.

\begin{figure}[!hbpt]
	\centering
		\includegraphics*[clip=true, trim=0mm 0.5mm 0mm 0.5mm,width=8cm]{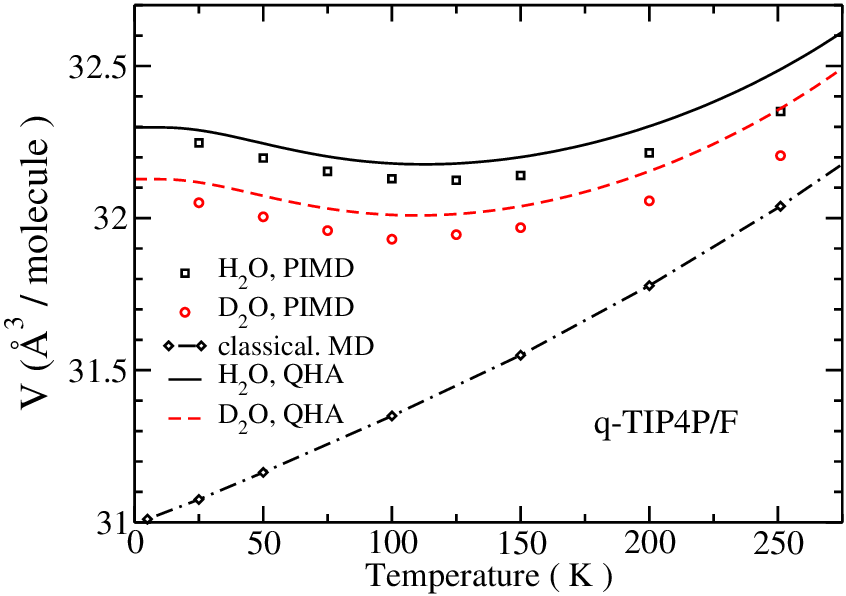}
	\caption{(Color online). Volume per molecule for different isotopes 
                 calculated using the q-TIP4P/F force field both with PIMD~\cite{Ramirez11} simulations and the QHA.}
	\label{fig:cla}
\end{figure}

Anharmonic effects, beyond the QHA, 
could change the $V(T)$ curves calculated with this method.
However, experimental Raman spectra of ice Ih, as a function of pressure and 
temperature~\cite{Johari78,Johari84,Wilkinson84} show a relatively 
temperature-insensitive stretching band, with 
$\partial\omega/\partial T < 0.6$~cm$^{-1}$K$^{-1}$.
Such a small anharmonicity should not modify the QHA results.

To test the validity of the QHA in ice Ih, Fig.~\ref{fig:cla} compares
the QHA result for $V(T)$ with our fully anharmonic PIMD
simulations~\cite{Ramirez11}, using the 
q-TIP4P/F force field~\cite{Manolopoulos09} in both cases.
Differences between the QHA and PIMD increase with temperature, 
 but their overall agreement is satisfactory.
In particular, with this force field both calculations predict a normal
isotope effect~\cite{Ramirez11}, which includes a lattice contraction
when H is replaced by D and a convergence of H and D volumes with
increasing T, contrary to the experimental result.
To complement these results, we have repeated the calculations using the
polarizable TTM3-F\cite{ttm3f,paesani09} potential.
This force field has recently been shown to outperform q-TIP4P/F when compared to experiments
in PIMD simulations of H$_2\text{}^{18}$O and D$_2\text{}^{18}$O.
Results provided in Table~\ref{table:vol_ordered_unitcell} and in the SI~\cite{SI},
 show that this polarizable force field also fails to reproduce the anomalous isotope effect.
However, as it will be discussed below, TTM3-F improves over q-TIP4P/F, displaying a stronger anti-correlation
effect.

 Our DFT  simulations are performed using the {\sc siesta} 
method~\cite{{SiestaPRBRC},{SiestaJPCM}}, with norm-conserving pseudopotentials
and numerical atomic basis sets for the valence electrons.
We use a basis of triple-$\zeta$ polarized orbitals, a longer and improved 
version of the basis set used in Ref.~\onlinecite{Jue11}.
Given the precision needed to obtain accurate Gr\"{u}neisen parameters, 
which involve numerical second derivatives of the forces, 
we have used highly converged grids for real-space and k-sampling integrations.
All our residual forces were smaller than 10$^{-3}$ eV/\AA.
Details of the parameters and convergence tests are provided 
in SI~\cite{SI}.

Hexagonal ice Ih is the most common form of ice, where oxygens occupy the hexagonal wurtzite lattice sites.
The two covalently-bonded protons have six possible orientations, but are constrained by Bernal-Fowler ``ice rules'' to have one proton per tetrahedral O-O bond.
This structure is characterized by proton disorder~\cite{Hirsch04}.
Ice XI is the proton-ordered phase of ice that is formed by a transition from hexagonal ice Ih at 72 K if catalyzed by KOH$^-$~\cite{Tajima82}.
 The lattice parameters are slightly different from those of ice Ih, with the
hexagonal symmetry weakly broken~\cite{Leadbetter84}.
We have performed calculations for three crystal structures:
(i) The Bernal-Fowler structure, with 12 molecules per unit cell, 
a net dipole moment along the c axis, and proton order 
consistent with the hexagonal symmetry.
(ii) a version of the ordered ice XI structure with 4 molecules 
per unit cell, constrained to be hexagonal, 
also with a net dipole moment along the c axis; and
(iii) a 96-molecule proton-disordered structure with no net dipole 
moment~\cite{Ramirez11}.
The oxygen lattice disorder observed in experiments~\cite{kush87}, is
of the same order of magnitude as what we observe in our PIMD simulations of
the proton disordered structure~\cite{rms,Ramirez11}.

%
As the results obtained for cells (i) and (ii) are very close,
 in the following we will present results for structures (ii) and (iii).
Table \ref{table:vol_ordered_unitcell} shows the $T=0$ volumes for several 
isotope combinations of H, D, $^{16}$O, and $^{18}$O, for all the DFT functionals~\cite{PBE,revPBE,Lundqvist04,SolervdW} 
 and force fields used in this study.
Also included in this table is a 32 beads PIMD result for a single unit cell 
($\Gamma$ sampling) using the PBE-DF, compared to a calculation with 
the QHA for an identical system. The simulation was done at $T=200$ K.

As the experimental study from  R\"{o}ttger {\it et al.} did not consider 
H$_2\text{}^{18}$O, we have performed high resolution X-ray diffraction experiments
of the three isotopes H$_2$O, D$_2$O and H$_2\text{}^{18}$O.
Results are presented in the table.
Experiments were performed with samples of water frozen in thin-walled glass capillaries, 
open at both ends to accommodate the volume expansion upon freezing,
or in flexible polymide tubes.
In all cases Si powder within the sample promoted nucleation of multiple crystals and
 provided an internal standard for precise determination of X-ray wavelength. 
 X-ray measurements were performed at a temperature of 100 and 220K K at the X16C high-resolution powder
 diffractometer at the National Synchrotron Light Source, using X-rays of nominal wavelength 0.7~\AA. 
 Diffraction peak widths were on the order of $0.02^\circ$ FWHM.
 Independent samples were prepared by slowly ($\sim 0.05$ K/sec) cooling and by quenching with liquid
 N$_2$. 
Lattice parameters were determined from measured positions of 21 ice diffraction peaks. 
Reported error bars encompass selected measurements at different temperatures.
A detailed experimental study including temperature dependence
will be published in a future contribution.

\begin{table*} [ht] \footnotesize
\caption{DFT volume (in \AA$^3$/molecule) for proton ordered (H-ordered)  and 
proton disordered (H-disordered) ice Ih for different isotopes, obtained 
with the quasiharmonic approximation (QHA) or path integral (PIMD) simulations. 
k-mesh is the effective number of {\bf k} points for sampling the 4-molecule hexagonal Brillouin zone in the phonon calculation
(one for $\Gamma$-sampling). IS(A-B)=$\frac{V(A)}{V(B)}-1$, is the relative isotope shift for the exchange of isotope A by B.
The exchange and correlation (XC) functionals are:
PBE~\cite{PBE}, vdW-DF$^{\rm{PBE}}$~\cite{Lundqvist04,Jue11}, 
revPBE~\cite{revPBE}, and vdW-DF$^{\rm{revPBE}}$~\cite{Lundqvist04}. The force fields (FF) are q-TIP4P/F\cite{Manolopoulos09} and TTM3-F\cite{ttm3f}
$V_{cla}$ is the volume for classical nuclei. Also shown are the experimental
results from ref~\cite{Kuhs94} and the ones obtained in this work. Note they are at different temperatures.}
\begin{center}
\begin{tabular} {c c c c c c c c c c c}
\hline
\hline
\cline{1-9}
T(K) &k-mesh & Structure & Method & XC/EFF & $V_{cla}$ & H$_2$O & D$_2$O & H$_2\text{}^{18}$O & IS(H-D) & IS($\text{}^{16}$O -$\text{}^{18}$O)\\
\hline 
200 &   1 & H-ordered    &PIMD   & PBE                    &       & 31.02 & 31.21 &       & $-0.61\%$ &  \\
200 &   1 & H-ordered    &QHA    & PBE                    & 30.6  & 30.00 & 30.16 & 29.98 & $-0.53\%$ & $+0.07\%$  \\
\hline
0   &   24 & H-disordered &QHA   & q-TIP4P/F &            30.98 & 32.30 & 32.13 & 32.24 & $+0.53\%$ & $+0.18\%$    \\
0   &   24 & H-disordered &QHA   & TTM3-F &               31.66 & 31.67 & 31.67 & 31.67 & $+0.002\%$ & $+0.002\%$ \\
0   &   24 & H-disordered &QHA   & PBE                    & 29.91 & 29.93 & 30.04 & 29.91 & $-0.35\%$ & $+0.07\%$ \\
\hline
0   &   1 & H-ordered    &QHA    & PBE                    & 29.98 & 29.91 & 30.05 & 29.89 & $-0.47\%$ & $+0.07\%$ \\
0   &   1 & H-ordered    &QHA    & vdW-DF$^{\rm{PBE}}$    & 30.88 & 31.01 & 31.10 & 30.98 & $-0.29\%$ & $+0.10\%$ \\
0   &   1 & H-ordered    &QHA    & revPBE   	          & 32.84 & 32.88 & 32.98 & 32.85 & $-0.30\%$ & $+0.09\%$ \\
0   &   1 & H-ordered    &QHA    & vdW-DF$^{\rm{revPBE}}$ & 33.45 & 33.73 & 33.76 & 33.70 & $-0.09\%$ & $+0.09\%$ \\
\hline
0   & 729 & H-ordered    &QHA    & PBE                    & 29.98 & 30.09 & 30.19 & 30.07 & $-0.33\%$ & $+0.07\%$ \\
0   & 729 & H-ordered    &QHA    & vdW-DF$^{\rm{PBE}}$    & 30.88 & 31.17 & 31.22 & 31.14 & $-0.16\%$ & $+0.10\%$ \\ 
0   & 729 & H-ordered    &QHA    & revPBE                 & 32.84 & 33.18 & 33.23 & 33.15 & $-0.15\%$ & $+0.09\%$ \\ 
0   & 729 & H-ordered    &QHA    & vdW-DF$^{\rm{revPBE}}$ & 33.45 & 33.95 & 33.94 & 33.92 & $+0.03\%$ & $+0.09\%$ \\   
\hline
10  &     & disordered    & Exp\cite{Kuhs94} &   &                & 32.054(5)  & 32.082(3) & & $-0.089(18)\%$ &           \\
100 &     & disordered    & Exp\cite{Kuhs94} &   &                & 32.047(3)  & 32.072(5) & & $-0.079(18)\%$ &           \\
100 &     & disordered    & Exp (this work)  &   &                & 32.079(4)  & 32.103(4) & 32.058(4)  &$-0.076(18)\%$ & $0.064(18)\%$   \\
220 &     & disordered    & Exp\cite{Kuhs94} &   &                & 32.368(4)  & 32.437(3) & & $-0.212(15) \%$ &           \\
220 &     & disordered    & Exp (this work)  &   &                & 32.367(4)  & 32.429(4) & 32.357(4) & $-0.191(17)\%$ & $0.032(17)\%$  \\
\hline
\hline
\end{tabular}
\end{center}
\label{table:vol_ordered_unitcell}
\end{table*}

Except for vdW-DF$^{\rm{revPBE}}$, all the other XC functionals predict an anomalous isotope effect at $T=0$
when H is replaced by D, in agreement with experiments.
However, the isotope effect has the normal sign for the O atom.
Our experiments confirm this result, with a 0.06$\%$ volume contraction
when $^{18}$O replaces $^{16}$O (T=100K).
The comparison of structures (ii) and (iii) shows that the results are largely
independent of the ordering of the protons.
Agreement with experiments improves as Brillouin zone integration 
is improved.
With respect to their generalized gradient approximation (GGA) counterparts, vdW-DFs soften the afore-mentioned
anticorrelation effect, reducing the magnitude of the H$\rightarrow$D isotope shift,
but have little effect on the O shift. 
When phonons from the full Brillouin zone are accounted, the vdW-DF$^{\rm{revPBE}}$ fails to predict
the isotope shift at $T=0$. 
However the anomalous shift for this functional is recovered at $T>100$ K \cite{SI}.
%
%
%
%
Overall, vdW-DF$^{\rm{PBE}}$ has proven to be very robust for a variety of structural and dynamical properties of water~\cite{Lundqvist04,mvfsprl06,Galli11}.
It also gives our best lattice constant for ice at $T$=0.
Therefore, in the following we use this functional and the QHA to explore the volume
of ice Ih in structure (ii), including full Brillouin zone phonon integration, 
as a function of isotope masses and temperature.
However, the results, and in particular the anomalous isotope effect, 
are very robust and largely independent of the XC functional, 
harmonic approximation, system size, and proton ordering.

\begin{figure}[ht]
	\centering
		\includegraphics[clip=true, trim=0mm 0.5mm 0mm 0.5mm,width=8cm]{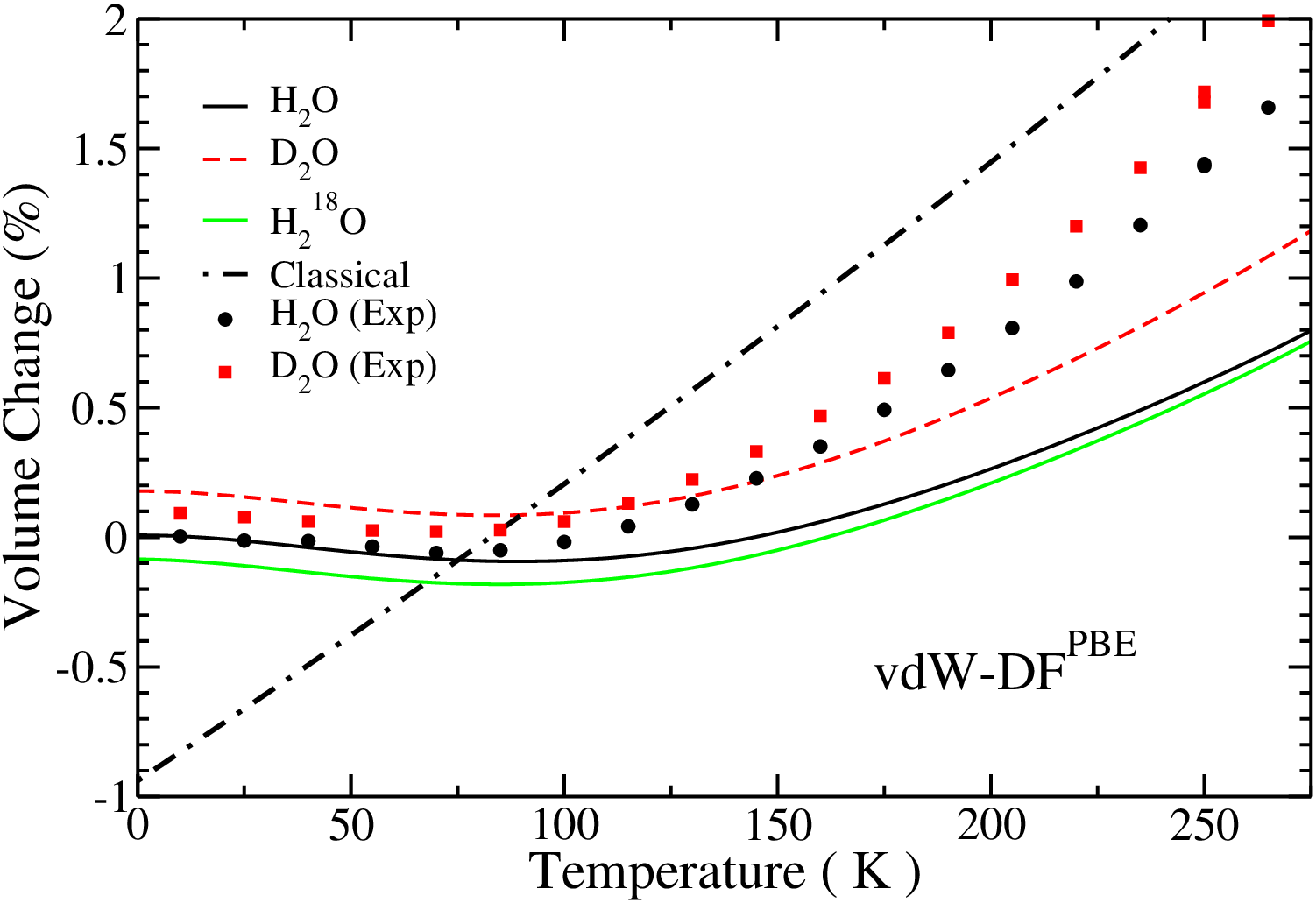}
	\caption{(Color Online). Volume change $V(T)/V_{\rm{H}_2\rm{O}}(0)-1$,
relative to H$_2$O at $T=0$, for different isotopes 
calculated using the QHA with the  
vdW-DF$^{\rm{PBE}}$ functional. 
Also shown are the experimental results 
from Ref.~\onlinecite{Kuhs94}.}
	\label{fig:a_T}
\end{figure}

Fig. \ref{fig:a_T} shows $V(T)$ of ice Ih for standard isotope substitutions
of H and O.
Experimentally, the anomalous H$\rightarrow$D isotope effect increases from 
$0.09\%$ at $T=10$ K to $0.25\%$ at $T=250$ K, 
and this increase is reproduced by our calculations 
(from $0.16\%$ to $0.32\%$).
The classical volume becomes larger than any of the quantum results 
above 100 K.
This implies that, for larger temperatures, a classical isobaric 
{\it ab initio} molecular dynamics simulation of ice will overestimate 
its volume, relative to a quantum PIMD simulation.

\begin{figure}[ht]

	\centering
		\includegraphics[clip=true, trim=0mm 0.5mm 0mm 0.5mm, scale=0.65]{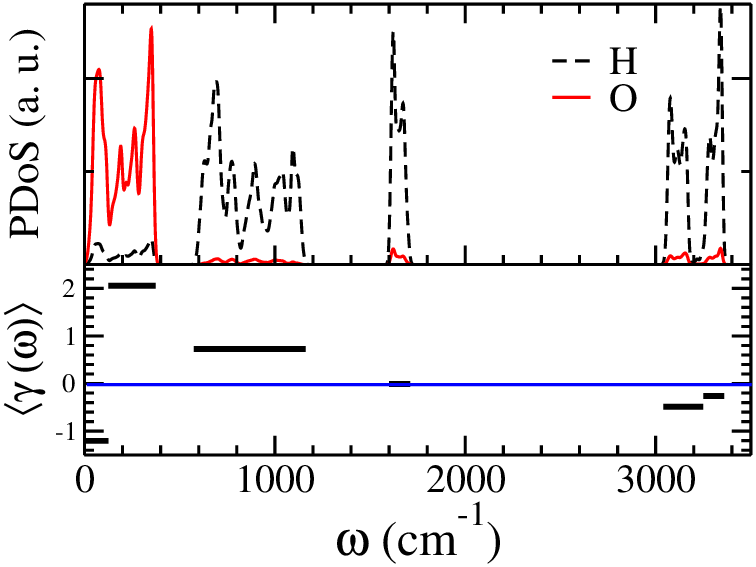}
	\caption{(Color online). 
Top: Density of vibrational states for H$_2$O, projected onto H and O atoms, for the 
ordered ice Ih structure, as obtained with vdW-DF$^{\rm{PBE}}$ functional. 
Bottom: average Gr\"{u}neisen constants of the different modes. 
}
	\label{fig:dos1}
\end{figure}

%
%
To analyze the complex isotope-dependent free energy surface,
we plot in figure \ref{fig:dos1} the vibrational density of states, 
projected onto the two atomic species, as well as the average 
Gr\"{u}neisen constants of the different phonon branches.
The volume change upon isotope substitution at $T=0$ 
is approximately proportional~\cite{Ziman,Allen94} to the average value 
$\langle \omega_k \gamma_k \rangle$.
Negative $\gamma_k$'s imply a softening of the modes with increasing pressure,
favoring smaller volumes for lighter isotopes.
Thus, the anomalous isotope effect, $V$(H$_2$O) $< V$(D$_2$O),
is due to the negative $\gamma_k$ of the covalent OH stretching modes
($\omega \sim$ 3100-3500 cm$^{-1}$)
which have more than 95$\%$ H weight.
Although the likewise H-dominated librational modes 
($\omega \sim$ 600-1000 cm$^{-1}$) 
have a positive $\gamma_k$, they are not enough to balance the 
volume shrinking contribution of the stretching frequencies.
In the case of the TTM3-F force field, the two contributions are very small
and almost completely cancel out, effectively producing
a very harmonic ice crystal, with a tiny anomalous effect at high $T$.
On the other hand, q-TIP4P/F largely underestimates the OH--O/OH anticorrelation, 
predicting a normal isotope effect for all temperature ranges.
Long range torsional modes, representing rigid tetrahedral rotations with
$\omega \approx 50$~cm$^{-1}$, also have $\gamma_k<0$.
These modes are responsible for the negative thermal expansion below 70 K.
The bending modes, with $\omega_k=1700$~cm$^{-1}$ and $\gamma_k \approx 0$ 
have little effect on volume.

Substitution of $^{16}$O by $^{18}$O affects mainly the low frequency modes, 
dominated by positive values of $\gamma_k$, producing a normal isotope effect.
The temperature dependence of the isotopic O substitution is also normal,
and the volume shift is 50\% smaller at $T$=220 K than at $T$=100 K.
Somewhat surprisingly, the net effect at $T=0$, relative to classical 
nuclei, is dominated by quantum oxygen, resulting in a quantum volume $\sim$1\% larger,
This small expansion (10 times smaller than that of Ne) 
is a consequence of two competing anharmonicities:
the contraction effect of H-dominated stretching modes and the
expansion effect of librational and translational modes.
%
%
As $T$ increases, the contribution of the stretching modes
becomes dominant, causing the net quantum effect to change sign and to become anomalous 
above $\sim$70 K.
This dominance increases with $T$, 
making the volume shift four times larger at the melting temperature than at $T$=0.
These results are not inconsistent with the requirement that, at high enough T
the isotope shift is isotope independent,
but we find that the convergence towards the classical limit
starts at T$>\sim$900 K.

Our results may also have significant implications for the understanding of nuclear 
quantum effects in liquid water (in which the anomalous isotope shift is experimentally larger than in ice~\cite{Kell67}).
PIMD simulations, 
using the q-TIP4P/F and TTM3-F EFFs, produce a less structured liquid than
classical MD simulations~\cite{Manolopoulos09,Ramirez2010a,ttm3f}.
However, as we have seen, these EFFs
do not reproduce the anomalous isotope effect in ice,
because they fail to describe accurately the derivatives of the frequencies,
which govern the anharmonicities and the nuclear 
quantum effects in the structure and dynamics.
This suggests that these models may be inadequate to reproduce
some quantum effects in the liquid as well as in the solid.
Therefore, the observed loss of structure in the liquid, for quantum vs classical nuclei,
should be reanalyzed with an
EFF that reproduces the anomalous quantum effects in ice.

In conclusion, we have shown that the anomalous nuclear quantum effects on the volume of
ice can be fully understood using the quasi-harmonic approximation with density functional theory.
The study fully explains a rare, seldom mentioned, property of ice which should
be included in the list of water anomalies~\cite{water-anomalies}, as an example 
in which quantum effects are anomalous and increase with temperature.

We thank Christian Thomsen for suggestions at the early stage of this work.
The work at Stony Brook University is supported by DOE award numbers DE-FG02-09ER16052  (MVFS)
and DE-FG02-08ER46550 (PBA). That at Madrid, by Spain's MCI grant FIS2009-12721-C04.
Use of the National Synchrotron Light Source, Brookhaven National Laboratory, 
was supported by the U.S. Department of Energy, Office of Basic Energy Sciences, under Contract No. DE-AC02-98CH10886.

\bibliography{PaperBetul}

\begin{thebibliography}{38}
\bibitem [{Gap()}]{GapIce}%
  \BibitemOpen
  \href@noop {} {}\bibinfo {note} {One can neglect thermal contributions to the
  electronic free energy because of the large band gap of ice.}\BibitemShut
  {Stop}%
\bibitem [{rms()}]{rms}%
  \BibitemOpen
  \href@noop {} {}\bibinfo {note} {It should be noted that a small amount of
  oxygen disorder, on the range of $~\pm 0.008$\AA\ is accounted for in our
  proton disordered structures at, and in the anisotropy of the ice XI lattice,
  at T=0. The O atom root mean square displacement (r.m.s) is even larger in
  the PIMD simulations (rms = $~\pm 0.07$\AA, T=100K; $~\pm 0.17$\AA, T=250K),
  which include a combination of both thermal motion and extra displacements
  due to the anisotropy associated to the proton disorder.}\BibitemShut {Stop}%
\end{thebibliography}

\end{document}